\documentclass[sigconf]{acmart}
\usepackage{hyperref}
\usepackage{url}
\usepackage{enumitem}
\usepackage{booktabs}
\usepackage{balance}
\AtBeginDocument{%
  \providecommand\BibTeX{{%
    \normalfont B\kern-0.5em{\scshape i\kern-0.25em b}\kern-0.8em\TeX}}}

\setcopyright{rightsretained}
\acmConference[SIGIR eCom'22]{ACM SIGIR Workshop on eCommerce}{July 15, 2022}{ Madrid, Spain}
\acmYear{2022}
\copyrightyear{2022}
\makeatletter
\renewcommand\@formatdoi[1]{\ignorespaces}
\makeatother
\acmISBN{}

\begin{document}

\title{Learning to Ask Critical Questions for Assisting Product Search}

\author{Zixuan Li}
\authornote{Zixuan Li is under the Industrial Postgraduate Program supported by Singapore Economic Development Board, Shopee Singapore Private Limited and National University of Singapore.}
\email{e0348832@u.nus.edu}
% \orcid{0003-0370-6538}
\affiliation{%
  \institution{Shopee Singapore Private Limited}
  \country{}
}
\affiliation{%
  \institution{Sea-NExT Joint Lab}
  \country{}
}
\affiliation{%
  \institution{National University of Singapore}
  \country{}
}

\author{Lizi Liao}
\email{lzliao@smu.edu.sg}
\affiliation{%
  \institution{Singapore Management University}
  \country{}
}

\author{Tat-Seng Chua}
\email{dcscts@nus.edu.sg}
\affiliation{%
  \institution{Sea-NExT Joint Lab}
  \country{}
}
\affiliation{%
  \institution{National University of Singapore}
  \country{}
}

\renewcommand{\shortauthors}{Li, et al.}

\begin{abstract}
Product search plays an essential role in eCommerce. It was treated as a special type of information retrieval problem. Most existing works make use of historical data to improve the search performance, which do not take the opportunity to ask for user’s current interest directly. Some session-aware methods take the user’s clicks within the session as implicit feedback, but it is still just a guess on user's preference. To address this problem, recent conversational or question-based search models interact with users directly for understanding the user's interest explicitly. However, most users do not have a clear picture on what to buy at the initial stage. Asking critical attributes that the user is looking for after they explored for a while should be a more efficient way to help them searching for the target items. In this paper, we propose a dual-learning model that hybrids the best from both implicit session feedback and proactively clarifying with users on the most critical questions. We first establish a novel \textit{utility} score to measure whether a clicked item provides useful information for finding the target. Then we develop the dual Selection Net and Ranking Net for choosing the critical questions and ranking the items. It innovatively links traditional click-stream data and text-based questions together. To verify our proposal, we did extensive experiments on a public dataset, and our model largely outperformed other state-of-the-art methods.
\end{abstract}





\maketitle

\section{Introduction}
\label{sec:intro}
Online shopping provides a convenient and cost-effective shopping method for people over the Internet, which has ignited an increasing interest in eCommerce technologies. Product search is one of the main traffic sources that drives conversion. Users usually formulate queries to express their needs and find products of interest by exploring the retrieved results. The majority of such methods apply some similarity functions to match the query and documents that describes a product. Under such cases, the query and product documents are usually represented as vectors in observed or latent vector space to compute the distances \cite{ai2017learning,guo2019attentive}. However, the search queries are often too general to capture the minute details of the exact product that the user is looking for \cite{zou2019learning}. 

Hence, there are two broad branches of research trying to address this problem. On one hand, session-aware recommendation makes use of the user's behavior data within the session to capture user's current interest, \textit{i.e.}, predicts the successive item(s) that an anonymous user is likely to interact with. Two major classes of approaches are applied: Markov-based models and DNN-based models \cite{meng2020incorporating}. The former emphasizes on capturing sequential patterns such as in the popular FPMC method \cite{rendle2012factorization}. Such models often emphasize the last click in history while ignoring previous behaviors. Hence, they often adopt static representations of user intentions, which may result in sub-optimal performance. 
The later one employs deep learning models such as RNNs to capture users' general preferences and current interests together. They leverage recurrent deep neural networks to encode historical interactions into hidden vector states, which leads to general and more informative representations. Nonetheless, there is still room for improvement since such models over-emphasize on monotonic behavior chains while failing to consider the more complex transition patterns among items.

On the other hand, interactive methods allow users to directly specify their needs in details. It is often realized under a conversational search and recommendation framework, where user preferences are clarified via aspect-value pairs \cite{zhang2018towards,Bi2019negativefeedback} or even unstructured text \cite{xiao2021endtoend}. 
There are also efforts prompting users with clarification questions under the traditional search setting with a query \cite{zou2019learning,zamani2020generating}. These methods largely enhance the accuracy of user preference modeling and flexibility of interaction. However, these conversational product search methods require intensive user involvement. They assume that the user has a clear target in mind, which does not hold true especially in early browsing stages \cite{hatt2020early}.

In this work, we propose \textbf{DualSI} -- a dual-learning model to hybrid the best from both session-aware recommendation and interactive methods. It makes use of the implicit session feedback to model user interests and proactively ask users the most critical questions to shorten the exploring session. Different from the traditional \textit{relevance} score widely used in product search, we first establish a new \textit{utility} score to measure how much useful information a clicked item provides for finding the target in a given context. Then we develop a Transformer-based selection module to calculate such contextualized \textit{utility} score which is able to capture the complex transition patterns beyond monotonic chain patterns. By further linking the most critical questions to the highest \textit{utility} scores, a dual ranking model predicts the target item based on the user responses. We conduct extensive experiments on a public dataset, and it shows that the proposed DualSI model largely outperforms baselines from both session-aware recommendation and interactive search.

To sum up, the main contributions of this work are as follows: 
\begin{enumerate}
\item We explicitly define a novel \textit{utility} score which measures whether a clicked item provides useful information in certain context for finding the target product.  
\item We design a dual-learning framework that incorporates a Transformer-based selection module for contextualized \textit{utility} score calculation and a dual ranking module for target item prediction. 
\item Extensive experiments demonstrate the effectiveness of our proposed method. Qualitative analysis results also show that the proposed method not only asks the right question for target finding but also helps to shorten the exploring session for users.
\end{enumerate}

\section{Related Work}
In this section, we provide a brief overview of the existing research efforts closely related to our work. Making use of user's behavior data within the session as implicit feedback for understanding user's current preferences is commonly known in session-aware recommendation. The idea has also been used in a few search related works. We will discuss them together under the session-aware recommendation subsection. They are highly relevant as our work also allows users to explore first and uses the within-session implicit feedback as input. Another important topic is interactive search. Our proposed method also learns to efficiently interact with users via question answering to improve search quality. Furthermore, we also provide an overview about works regarding learning to ask, and briefly discuss about their connections to our work.

\subsection{Session-aware Recommendation}
Session-aware recommendation predicts the successive item(s) that an anonymous user is likely to interact with, according to the implicit feedback within the session\cite{Wu2017SessionawareIE,10.1145/3442381.3450005,NSAR}. Generally speaking, there are two major classes of approaches to leverage sequential information from users’ historical records: Markov-based models and DNN-based models. In early days, Markov chains are widely applied to capture sequential patterns between consecutive user-item interactions \cite{shani2005mdp}. For example, it was applied to characterize users’ latest preferences with the last click, but the previous clicks and the useful information in the long sequence are neglected \cite{sarwar2001item}. Rendel \textit{et al.} proposed a hybrid model FPMC \cite{rendle2012factorization}, which combined Matrix Factorization and Markov Chain to model sequential behaviors for next basket recommendation. However, the adoption of static representations for user intentions is a major problem of such methods. 

More recently, the DNN-based models such as Recurrent Neural Networks (RNNs) have been employed in session-aware recommendation and demonstrated state-of-the-art performance. They are widely used to capture users’ general interests and current interests together through encoding historical interactions into a hidden state (vector). The method in \cite{Hidasi2016SessionbasedRW} is among the pioneer works that proposed a deep RNN-based model to encode items sequences for recommendation. Jing \textit{et al.} \cite{li2017neural} further improved it through adding extra mechanism to tackle the short memory problem inherent in RNNs. In addition, to explicitly model the effects of users’ current actions on their next moves, the model in \cite{liu2018stamp} utilized an attention net to model user’s general states and current states separately. Besides, Hidasi \textit{et al.} proposed two variations to improve their model performance through adjust loss functions \cite{hidasi2018recurrent,hidasi2016parallel}. There are also works such as \cite{wu2019session} adopting Graph Neural Networks (GNNs) to capture the complex transition patterns among the items in a session rather than the transition pattern of single way. Beside recommendation, some search work also incorporate the session-aware concept. Bi \textit{et al.} makes use of implicit feedback for re-ranking the search results in the following page \cite{bi2020leverage}. However, all these methods rely on inferred user interests or preferences which might be wrong or inaccurate. In this paper, we allow the model to solicit user feedback explicitly via asking critical questions.

\begin{figure}[t]
	\centering
	\includegraphics[width=0.45\textwidth]{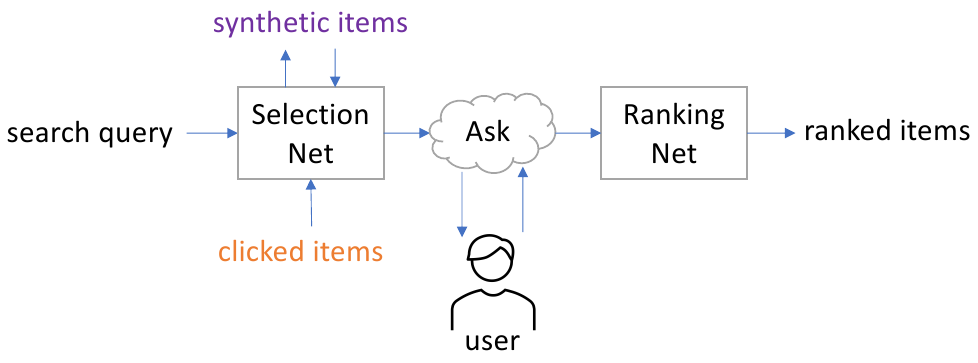}
	\caption{The DualSI framework. The Selection Net calculates contextualized utility scores for clicked items, which helps to select the most critical questions. The dual Ranking Net learns to predict the target item based on user's feedback.}
	\vspace{-0.3cm}
	\label{fig:framework}
\end{figure}

\begin{figure*}
\centering
\vspace{+0.2cm}
\includegraphics[width=0.96\textwidth]{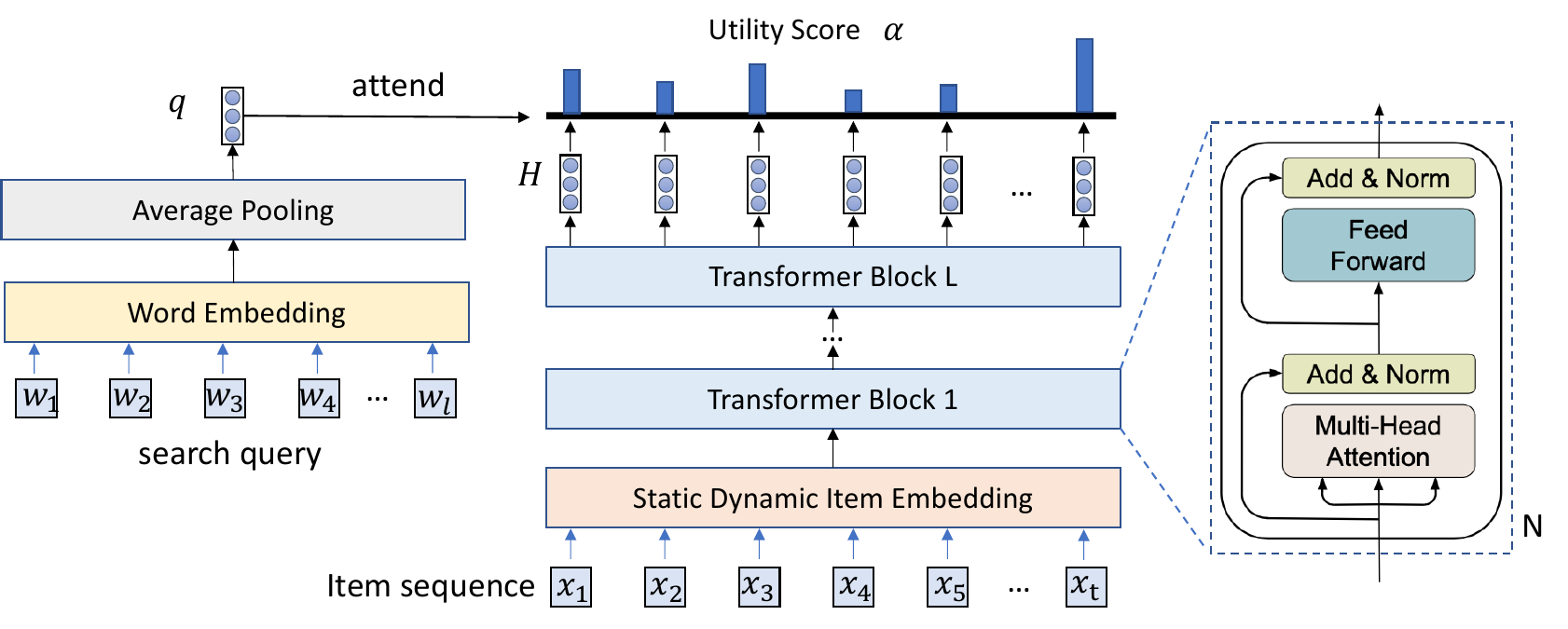}
\caption{The Selection Net architecture. The item sequence obtains contextualized representations via static dynamic item embeddings and self-attention blocks. The query signals attend over them for better utility score calculation. Details of the Static Dynamic Item Embedding module are further provided in Figure \ref{fig:static_dynamic}.}
% \vspace{-0.3cm}
\label{fig:selectionNet}
\end{figure*}

\subsection{Interactive Search}
Interactive search puts human in the loop whereby user's preferences of a specific query can be directly captured \cite{christa_towards,yueming_2018,EAR}. Zhang \textit{et al.} proposed a conversational search and recommendation framework and implemented it using a Multi-Memory Network architecture \cite{zhang2018towards}. In this work, the product search is done after clarifying user's preferences on different aspect-value pairs via dialogue. Considering the imperfection of product attribute schema which has been commonly used for questions generation in conversational search \cite{zhang2018towards,Bi2019negativefeedback}, Xiao \textit{et al.} introduced an end-to-end conversational product search system that also incorporates unstructured text to enhance the product knowledge \cite{xiao2021endtoend}. Other than purely focusing on the textual modality, there are also works incorporating images into the dialogue \cite{10.1145/3240508.3240605,moon-etal-2020-situated}. However, all these conversational product search approaches require intensive user involvement and assume the users already have specific item in mind which might not be true. Shoppers usually do not have a clear picture of the item in mind when they are at early shopping phase \cite{chen2009information,hatt2020early}.

Another branch of work for interactive search prompts users with clarification questions in the traditional search setting \cite{zou2019learning,zamani2020generating}. After user issuing the query, the clarification question is asked with options provided for the user to interact with. Zou \textit{et al.} introduced a Bayesian-based product search method which sequentially prompts users on their expected entities in the target product document \cite{zou2019learning}. A fixed number of questions are asked one by one whereby the next question is generated based on the user's answers on previous questions. However, their proposed algorithm can only handle seen queries. Similarly, \cite{Zou_2020_Qrec} makes use of historical user ratings for offline initialization and learns to ask a sequence of questions so as to understand user preferences for recommending items. Other than text query, in \cite{skopal2018image}, it asks user to click on his/her interested product query image region, and use the information for generating more accurate results that targets the user's specific interest. These work also have similar problems as mentioned for conversational search above. Our work avoids these issues by only asking the most critical questions on top of the observed user click behaviors.

\subsection{Learning to Ask}
Learning to ask aims to seek enough information for specific tasks with as few questions as possible. Generally speaking, it is naturally involved in all conversational and user interaction related works. Some papers studied this topic specifically in a 20 questions game setting, whereby one party think of an object and the other party tries to guess by asking no more than 20 questions. For example, Wu \textit{et al.} introduced an entropy-based ranking algorithm to calculate the object-question relevance matrix utilizing user's answer distributions \cite{Wu_2018}. Almost in the same time period, in \cite{Chen_2018}, the authors used deep reinforcement learning and general matrix factorization for the agent to learn smart questioning strategies. Similarly, Hu \textit{et al.} proposed a policy-based reinforcement learning method that enables the agent to learn optimal policies via continuous interaction with user \cite{Hu2018Playing2Q}.
However, these works require heavy simulation for the agent to learn well, which has been avoided in our work by leveraging user's implicit feedback.

\section{Method}
This section provides a detailed description of the proposed
method. The overall framework is shown in Figure \ref{fig:framework}. Our approach consists of two parts: (a) the calculation of utility score using Selection Net for critical question selection; and (b) the product ranking using the dual Ranking Net given user's within session behavior and answers to the critical questions. This design leverages both user's implicit feedback and explicit question answering, which is inspired by the fact that shoppers usually gets clearer on the exact item to purchase after some initial browsing, and directly clarifying with the user afterwards on the selected critical questions would efficiently improve the shopping experience. The detailed training scheme is also presented in the end of this section.

\subsection{Contextualized Selection Net}
The core of our method is the contextualized Selection Net. It takes the click-stream data as input for context capturing and question selection via the utility score.
\subsubsection{Utility Score} \hfill\\
Search queries are usually very general, especially when a user does not have a clear picture on what to buy or cannot describe precisely the item he/she is looking for. After issuing the search query, a user usually needs to browse for a while on different products to eventually find the `right' item. During the browsing process, the user's idea on what exactly to buy becomes clearer. Each clicked item in the click sequence carries different level of information on helping the user to find the target item. We explicitly measure it by introducing the utility score $\alpha$. Intuitively, higher utility score implies that the item carries more critical information related to the target product within its search context. 

In our implementation, the utility score is instantiated by calculating the attention score between the search query \(q\) and contextualized hidden states of the clicked items $H=[h_{1},h_{2},...,h_{t}]$ as indicated in Figure \ref{fig:selectionNet}. We design the loss function to enable the training of the Selection Net to learn to generate higher scores for clicked items that carries more information leading to the target. We hypothesis that such `information' is expressed by the item attributes, whereby each product can be uniquely described using a set of attributes. Such `information' is also reflected in the relative occurrence times of attributes inside the click stream. This is intuitive as these repeatedly clicked items/attributes are usually more interested by the user. In the inference stage, the synthetic items mentioned in Figure \ref{fig:framework} are generated using different combinations of attributes. By inputting each of them together with the clicked items through the Selection Net, the synthetic item with highest utility score is chosen. Its attributes are used as critical questions to clarify with the user, and the hidden states \(H'\) after the last attention layer are then used in the ranking stage. More design and implementation details are covered in the next subsection.

\subsubsection{Contextualized Selection} \hfill\\
With the concept of utility score being introduced, we now explain how the Selection Net is designed. As shown in Figure \ref{fig:selectionNet}, to capture the complex search and browsing context, we encode the clicked items with static dynamic item embedding layer first, and then pass through a transformer encoder which can access information of all clicked items in the sequence to obtain the contextualized representations. The search query \(q\) then attend on the output of the transformer module (i.e. hidden states \(H\)). More attention should be given to the items with higher utility score. Hence, we use the attention score directly. The ground truth score for the target item is 1, as the target item is ideally the one carries the highest utility score. The scores of other items are suppressed.

{\bf Query Embedding.} As suggested in \cite{bi2020leverage}, more complex encoding approaches are not expected to have a better performance than the simple average. A query issued by an user typically consists of multiple words, where the word embedding $\varepsilon(w) \in \mathbb{R}^{d}$ and \(d\) is the embedding dimension. In our implementation, we use the widely applied GloVe word embeddings \cite{pennington2014glove}. We average the word embeddings to obtain the corresponding query representation, i.e.,
\[ q=\frac{\sum_{w\in query}^{} \varepsilon(w)}{l} \tag{1}\]
where \(l\) is the number of words in the query.

\begin{figure}
\centering
\includegraphics[width=0.45\textwidth]{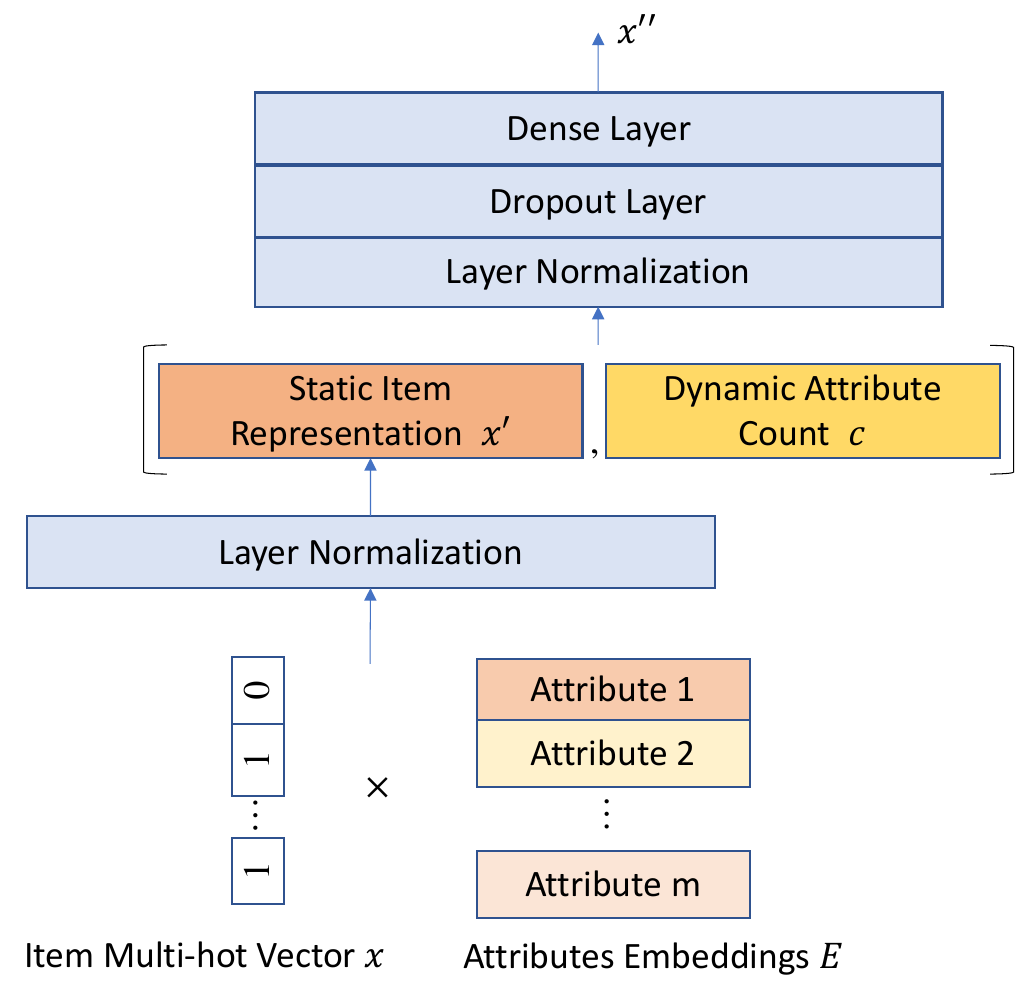}
% \vspace{-0.5cm}
\caption{The static item embedding concatenate with the dynamic attribute count feature to form a comprehensive representation for each item.}
% \vspace{-0.3cm}
\label{fig:static_dynamic}
\end{figure}

% {\bf Entity Embedding} Entities of a product can be extracted from its product title and other available text knowledge. Similar to query embedding, each entity \(e\) consists of multiple words \(w\). The entity embedding is the average of its word embeddings, i.e.,
% \[ \varepsilon(e)=\frac{\sum_{w\in e}^{} \varepsilon(w)}{\left| e \right|} \tag{2}\]
% where $\left| e \right|$ is the number of words in the entity.

% =[{0, ..., a_{e1}, ..., a_{ek}, ..., 0}]$, where $a_{e}=1$ for $e = e_{1}, ..., e_{k}$ and 0 everywhere else, where $e_{1}, ..., e_{k}$ index the corresponding attributes.
{\bf Static Dynamic Item Embedding.} Each item can be uniquely represented by the set of attributes associated with it. We firstly create a multi-hot vector $x \in \mathbb{R}^{m \times 1}$ for each item, where \(m\) refers to the number of distinct attributes of the product pool. With the corresponding attribute positions flagged as 1 in $x$, the static dense representation for an item is
\[ x' = LN(x^{T}E) \tag{2}\]
where $E=
\begin{bmatrix}
e_{1}\\
...\\
e_{m}
\end{bmatrix}
$ and $e_{m}$ is the embedding for the \(m-\)th attribute generated in a similar way as query embedding. $LN(\cdot)$ refers to the layer normalization operation.

On top of the static item embedding mentioned above, we concatenate it with a dynamic feature called Attribute Count \(c\) before inputting it to the transformer encoders:
\[ x'' = Dense(Dropout(LN([x' , c]))) \tag{3}\]
where $Dropout(\cdot)$ refers to the dropout layer and $Dense(\cdot)$ refers to the fully-connected feed-forward layer.
It is intuitive to assume that attributes appeared more frequently in the clicked item sequence are more relevant to the target item. This is also verified by looking at its capability in finding target items in our preliminary analysis, hence it is used here explicitly. The Attribute Count feature is a simple average of the accumulated appearance counts for each flagged attribute associated with the item by the time it is clicked.  After the concatenation, a dynamic representation is obtained with dimension \(d+1\). To normalize across the features, layer normalization is used. In addition, a dropout layer and a dense layer is applied which projects it to dimension \(d\). An illustration on how the dynamic representation is obtained for each input item can be found in Figure \ref{fig:static_dynamic}. We will also show how the popular attributes affect the learning-to-ask performance in the ablation study later.

\begin{figure*}
\centering
\includegraphics[width=0.94\textwidth]{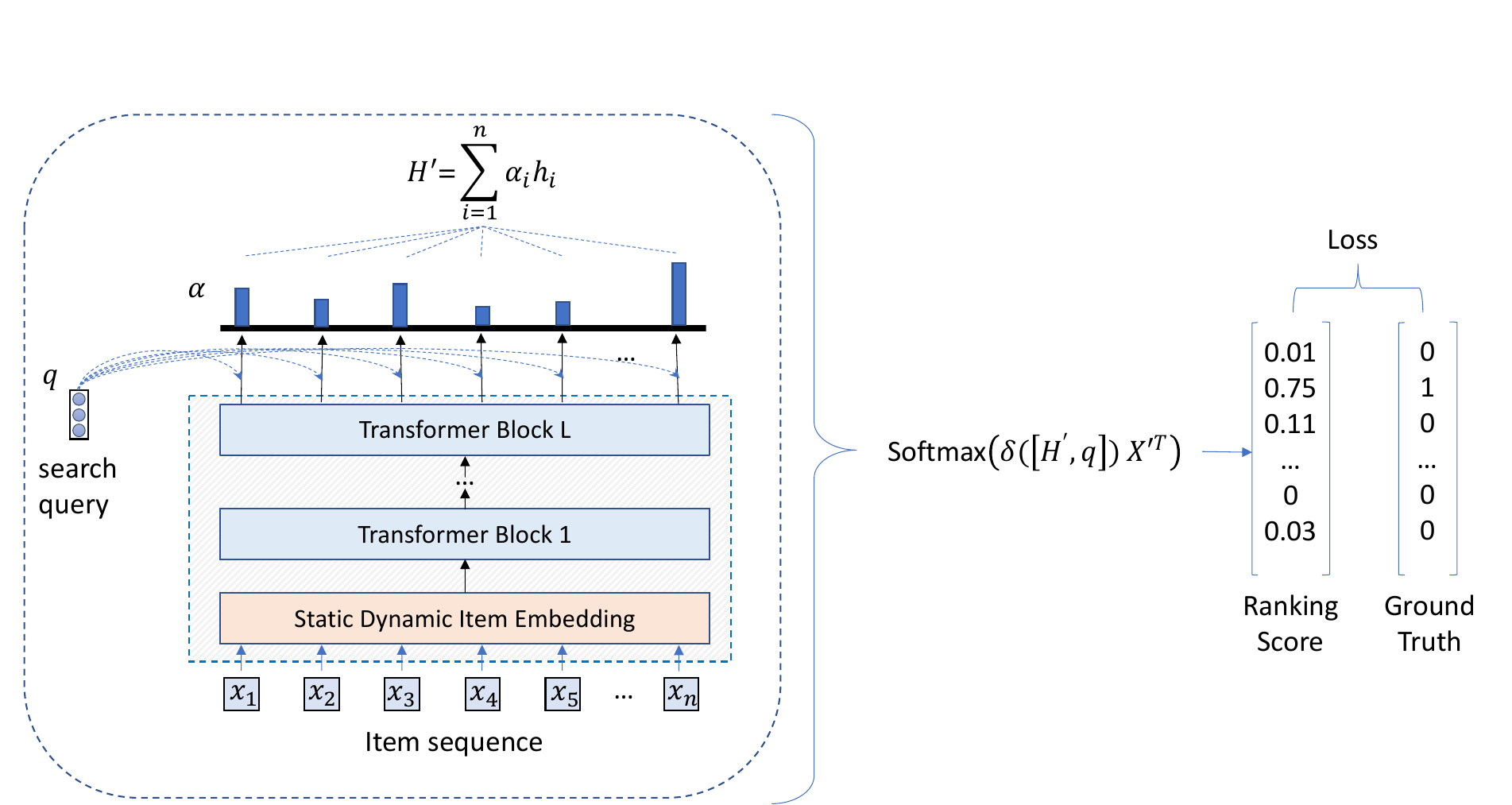}
% \vspace{-0.5cm}
\caption{The Dual Ranking Net. There are normalization layer, flatten layer, dropout layer and dense layer used in the $\delta$ network. After multiplying with the item pool \(X'^{T}\), there are again a normalization layer, some dense layers and dropout layers. They are omitted here for a concise and clear illustration on the main concept.}
% \vspace{-0.3cm}
\label{fig:rankingNet}
\end{figure*}

% ${i_{1:t}}$, where ${i_{1:t}} \in \mathbb{R}^{t\times (d+1)}$
{\bf Contextualized Representation.} Now, we map the comprehensive representation of the sequence of clicked items to contextualized embeddings. Specifically, we map these into $H=[h_{1},h_{2},...,h_{t}]$ using \(L\) transformer encoder layers following the definition in \cite{NIPS2017_3f5ee243}. We adopt the transformer module because of its capability in capturing the complex context information among the clicked items. In the actual implementation, we set L=1 as the experiment dataset is not too large to model. It can be adjusted accordingly for other use cases. The queries \(Q\), keys \(K\) and values \(V\) are all referring to the linear transformations of $x''_{1:t}$ in the case of self-attention in transformer encoder

\[H=TransformerEncoder(x''_{1:t})\tag{4}\]
Inside the transformer encoder, we use four heads for multi-head attention. 
\begin{align*}
\tag{4.1} MultiHead(x''_{1:t})=[head_1,head_2,head_3,head_4]W^O\\ 
\tag{4.2} head_i= Attention(x''_{1:t}W_i^{Q}, x''_{1:t}W_i^{K},x''_{1:t}W_i^{V})\\ 
\tag{4.3} Attention(Q,K,V) =  Softmax( \frac{(Q \cdot K^T)} {\sqrt{d}}) \cdot V 
\end{align*}
where $W^O$, $W_i^Q$, $W_i^K$ and $W_i^V$ are the projection parameter matrices.
It is then followed by residual connections, layer normalizations and dense layers as in the original transformer paper \cite{NIPS2017_3f5ee243}, except we set N=1 for relatively small data size.

{\bf Utility Score.} The attention layer calculates the utility score for each item with contextualized $H$ and the query $q$:
\[\alpha_{i}= \frac{exp(q\cdot h_{i}^{T})}{\sum_{j=1}^{t}exp(q\cdot h_{j}^{T})} \tag{5}\]

{\bf Loss Function.} During training, the utility score $\alpha$ is maximized at the target item position by minimizing the mean absolute error (MAE) loss, i.e.,
\[\mathcal{L}_{selection}=\frac{\sum_{j=1}^{t}\left| \alpha_{j}-y_{j} \right|}{t} \tag{6}\]
where $y_{j}$ is the ground truth label at position \(j\).

The training of the Selection Net aims to gain the capability of identifying the item with the highest utility score, i.e. the target item. During testing, given that \(k\) attributes are to be asked at one question prompt, each \(k\)-attributes combination is encoded into the multi-hot vector to form a synthetic item $x_{synthetic}$, and then pass through the trained Selection Net together with the clicked items, i.e. $x_{1:t-1}$. The one with the highest utility score in the given click sequence is chosen to prompt to the user for feedback.

\subsection{The Dual Ranking Net}
With the given user behavior context and answers to the questions, the dual Ranking Net learns to rank the items. It reuses the parameters from the trained Selection Net as shown in Figure \ref{fig:rankingNet} (the shaded encoding part is frozen during training). After encoding the item sequence, we obtain the contextualized embedding \(H'\) from the encoder net
\[H'=\sum_{i=1}^{n}\alpha_{i} h_{i} \tag{7}\]
where \(n\) is the item sequence length including the synthetic item from selected questions. \(n\) equals to \(t\) only when the system prompts user questions one step before the target.

For the product relevance ranking, \(H'\) is firstly concatenated with the search query \(q\). Then the combined representation goes through a $\delta$ network which consists of a normalization layer, a flatten layer, a dense layer and a dropout layer. Then it is multiplied with the item pool \(X'^{T}\), where $X'=[x_{1}',x_{2}',...,x_{p}']$ and $p$ is the number of products. On top of that, it further connects to a normalization layer and a  two-layer feed-forward network. A softmax function is applied as the last layer to get the ranking score of each item.
\[score(X')=Softmax(\delta([H',q])X'^{T}) \tag{8}\]

 The cross entropy loss $\mathcal{L}_{ranking}$ is calculated between the predicted scores and ground truth $y'$ for the Ranking Net training.
\[\mathcal{L}_{ranking}=Cross\-Entropy(score(X'),y') \tag{9}\]

\subsection{Training and Inference Scheme}
Below is a detailed explanation on how the training and inference are done with the Selection Net and Ranking Net. 
\subsubsection{Training Stage} 
\begin{figure}
\centering
\includegraphics[width=0.45\textwidth]{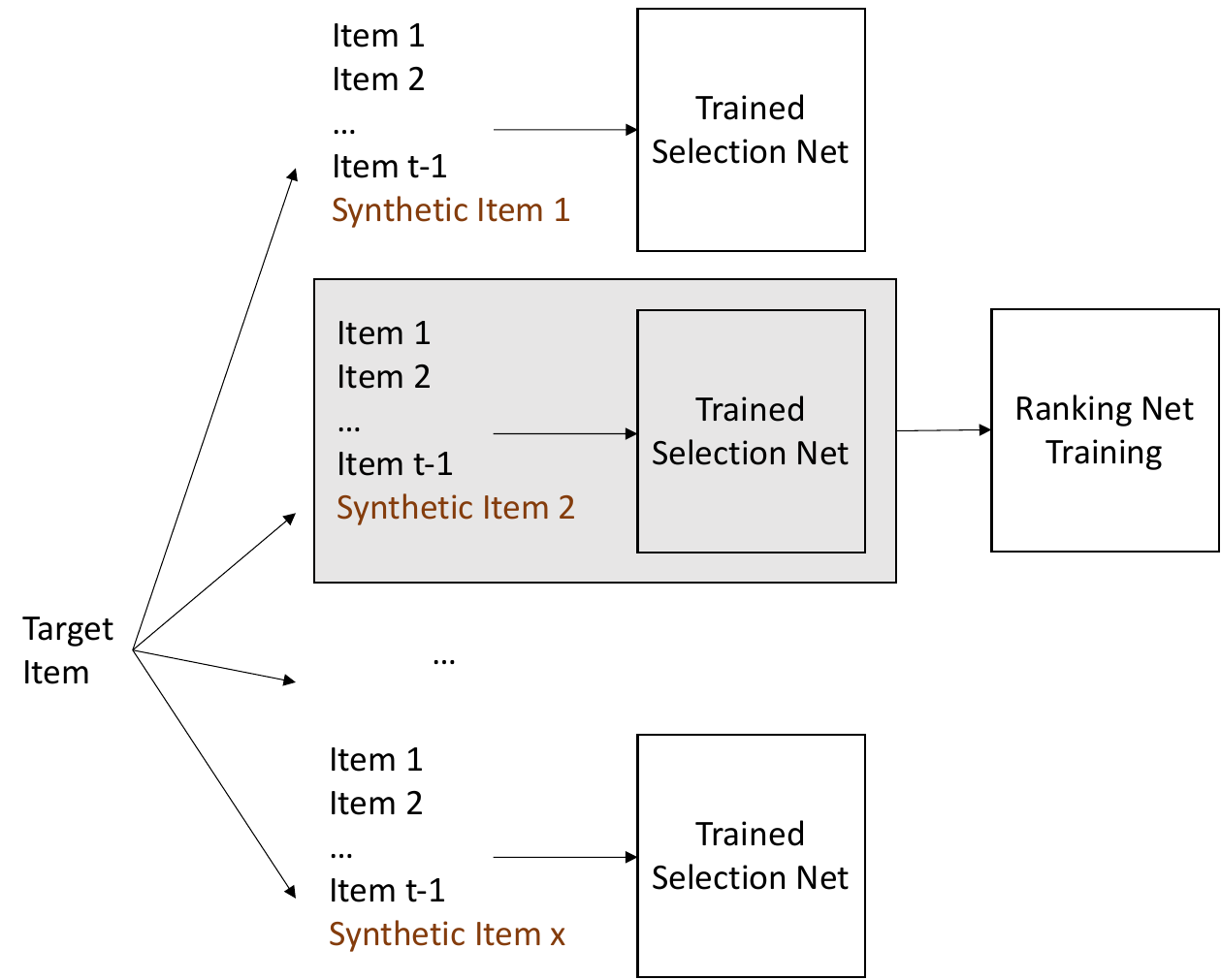}
% \vspace{-0.5cm}
\caption{Training scheme for the Ranking Net. The one in the grey box is chosen as the input for Ranking Net Training if synthetic item 2 has the highest score among all synthetic items generated based on the target item.}
% \vspace{-0.3cm}
\label{fig:rankingNetTraining}
\end{figure}

\hfill\\
For Selection Net, it takes the sequence of clicked items up to the target item for training as shown in Figure \ref{fig:selectionNet}. When it comes to Ranking Net, the item sequence is formed by the clicked items $x_{1:t-1}$ together with each of the synthetic items generated using the attributes appeared in the target item. After getting the utility score for each composed sequence, only the one with maximum utility score at the synthetic item position is passed through to the rest of the Ranking Net for training. An illustration can be found in Figure \ref{fig:rankingNetTraining}.  Padding and Masking techniques are used to standardize the input for different click sequence length.
\subsubsection{Inference Stage} \hfill\\
During testing stage, instead of generating synthetic items from target item attributes, they are generated from all possible combinations of the attribute pool, as it is for the Selection Net to decide on which question to ask. Each of these synthetic item combines with the n clicked items, and input into the trained Selection Net. The one with highest utility score is then used for generating the critical questions. If the attribute asked exist in the target item, it simulates that the user answered with `Yes' to this attribute. Based on the answer, the Ranking Net then takes the corresponding \(H'\) for relevance ranking. For cases where the user answered with `No' or answered `Yes' to some of the attributes only, we gets the input sequence updated accordingly for getting the \(H'\) for eventual ranking. 

\begin{table*}[t]
\centering
\renewcommand*{\arraystretch}{1.3}
\caption{\label{table:mainResult}Performance comparison regarding metrics such as MRR, Top-3 Accuracy and NDCG among baselines and variants of our model.}
\begin{tabular}{l|l|l|l|r|r|r|r|r|r} 
\hline
              & \multicolumn{3}{c|}{No Question Asking}                                                                                & \multicolumn{3}{c|}{Asking 1 Question/Attribute}                                                                                        & \multicolumn{3}{c}{Asking 2 Questions/Attributes}                                                                                      \\ 
\cline{2-10}
              & MRR~$\uparrow$                          & \begin{tabular}[c]{@{}l@{}}Top-3 \\Accuracy~$\uparrow$ \end{tabular} & NDCG~$\uparrow$                          & \multicolumn{1}{l|}{MRR~$\uparrow$ } & \multicolumn{1}{l|}{\begin{tabular}[c]{@{}l@{}}Top-3 \\Accuracy~$\uparrow$ \end{tabular}} & \multicolumn{1}{l|}{NDCG~$\uparrow$ } & \multicolumn{1}{l|}{MRR~$\uparrow$ } & \multicolumn{1}{l|}{\begin{tabular}[c]{@{}l@{}}Top-3 \\Accuracy~$\uparrow$ \end{tabular}} & \multicolumn{1}{l}{NDCG~$\uparrow$ }  \\ 
\hline
SR-GNN          & \multicolumn{1}{r|}{0.08606} & \multicolumn{1}{r|}{0.09105}                             & \multicolumn{1}{r|}{0.19237} & \multicolumn{1}{l|}{-}   & \multicolumn{1}{l|}{-}                                                        & \multicolumn{1}{l|}{-}    & \multicolumn{1}{l|}{-}   & \multicolumn{1}{l|}{-}                                                        & \multicolumn{1}{l}{-}     \\ 
\hline
SCEM          & \multicolumn{1}{r|}{0.16187} & \multicolumn{1}{r|}{0.18681}                             & \multicolumn{1}{r|}{0.27548} & \multicolumn{1}{l|}{-}   & \multicolumn{1}{l|}{-}                                                        & \multicolumn{1}{l|}{-}    & \multicolumn{1}{l|}{-}   & \multicolumn{1}{l|}{-}                                                        & \multicolumn{1}{l}{-}     \\ 
\hline
QSBPS         & -                            & -                                                        & -                            & 0.15971                  & 0.16169                                                                       & 0.25736                   & 0.18476                  & 0.20251                                                                       & 0.28644                   \\ 
\hline
Qrec         & -                            & -                                                        & -                            & 0.16746                  & 0.16908                                                                       & 0.25596                   & 0.17567                  & 0.17713                                                                       & 0.26595                   \\ 
\hline
DualSI\_randomQ & -                            & -                                                        & -                            & 0.16015                  & 0.17268                                                                       & 0.27312                   & 0.16015                  & 0.17268                                                                       & 0.27314                   \\ 
\hline
DualSI\_randomA & -                            & -                                                        & -                            & 0.19315                  & 0.21821                                                                       & 0.30364                   & 0.17613                  & 0.19623                                                                       & 0.28722                   \\ 
\hline
DualSI\_popular & -                            & -                                                        & -                            & 0.22443                  & 0.25275                                                                       & 0.33162                   & 0.24273                  & 0.27630                                                                        & 0.34705                   \\ 
\hline
DualSI          & \multicolumn{1}{r|}{0.16015} & \multicolumn{1}{r|}{0.17268}                             & \multicolumn{1}{r|}{0.27313}                            & \textbf{0.23310}          & \textbf{0.25746}                                                              & \textbf{0.33909}          & \textbf{0.24751}         & \textbf{0.28571}                                                              & \textbf{0.35075}          \\
\hline
% DualSI\_finetune          &0.09319                             &0.09733                                                         &0.20761                             & \textbf{0.27218}          & \textbf{0.30926}                                                              & \textbf{0.37691}          & \textbf{0.27038}         & \textbf{0.29827}                                                              & \textbf{0.37555}          \\
% \hline
\end{tabular}
\end{table*}

\begin{table*}[t]
\centering
\renewcommand*{\arraystretch}{1.3}
\caption{\label{table:ablation}Performance comparison of the proposed DualSI method among the settings of asking question 1-5 item clicks before the actual target item is clicked.}
\begin{tabular}{l|l|c|l|l|c|l} 
\hline
                      & \multicolumn{3}{l|}{Asking 1 Question/Attribute} & \multicolumn{3}{l}{Asking 2 Questions/Attributes}  \\ 
\cline{2-7}
                      & MRR~$\uparrow$      & Top-3 Accuracy~$\uparrow$  & NDCG~$\uparrow$                & MRR~$\uparrow$       & Top-3 Accuracy~$\uparrow$  & NDCG~$\uparrow$                  \\ 
\hline
1 click & 0.23310 & 0.25746        & 0.33909            & 0.24751  & 0.28571        & 0.35075              \\ 
\hline
2 clicks& 0.21790 & 0.24333        & 0.32586            & 0.23178  & 0.25432        & 0.33736              \\ 
\hline
3 clicks & 0.19721 & 0.22449        & 0.30584            & 0.19778  & 0.21350        & 0.30801              \\ 
\hline
4 clicks & 0.15445 & 0.16797        & 0.26799            & 0.16688  & 0.18838        & 0.28080              \\ 
\hline
5 clicks& 0.13399 & 0.14443        & 0.24608            & 0.14575 & 0.16484        & 0.25755              \\
\hline
\end{tabular}
\end{table*}

\section{Experiments}
In this section, we conduct extensive experiments mainly to answer the following research questions: 
\begin{itemize}
\item[\bfseries RQ1] Our proposed method leverages both implicit feedback and explicit feedback. Is it performing better than those using only one of them?  
\item[\bfseries RQ2] Does our method ask the right questions? Are the answers from users play an important role?  
\item[\bfseries RQ3] Would the proposed method manage to shorten the search sessions for user?
\end{itemize}

\subsection{Experimental Setup}
\subsubsection{Dataset} \hfill\\
We use the Diginetica dataset\footnote{http://cikm2016.cs.iupui.edu/cikm-cup} which was first introduced in 2016 for CIKM Cup on product search. It consists of query-full and query-less search related data. Query-full search refers to those triggered by an user issued query, while query-less search refers to those triggered by a category click. We only used the query-less search data as the amount of query-full search data is very small after processing. For query-less search data we are using, the category ID is used to represent the search query, which is a common practice in many product search works \cite{Rowley2000ProductSI, zou2019learning}. We also filter out search sessions with less than 5 or more than 20 clicks before purchase. In principle, our DualSI can handle any sequence length. This preprocessing is mainly for a standard evaluation comparison later. After random split, 75\% of the data is used for training, 10\% for validation and 15\% as test set. To the best of our knowledge, this is the only public dataset that matches our setting.

\subsubsection{Question Pool Construction} \hfill\\
The question pool can be constructed by extracting attributes from product documents, which includes but not limit to product title, product description, product price range and etc. There are different methods for extracting product attributes from its document, such as using the TAGME tool \cite{tagMe}. In the Diginetica dataset we use, since only hashed product title and price are available, we took all the available words in these as possible attributes. Due to data limitation, the minimum number of attributes for each item is two. This is why we only did experiment on asking up to two questions. Nevertheless, this aligns with our aim of only asking critical questions without bothering the user too much. In addition, we perform simple pre-processing such as filtering out attributes that only appeared once.

\subsubsection{Evaluation Measures} \hfill\\
To evaluate our method, we use Mean Reciprocal Rank (MRR), Top-3 Accuracy
and Normalized Discounted Cumulative Gain (NDCG) as the evaluation metrics. In our setting, there is only one target product for each purchase, which makes MRR a proper metric to use to calculate the inversed rank value of the target item. The Top-3 Accuracy suggests whether the target item appears in the top-3 ranked items. NDCG is another important metric that is commonly used for product ranking performance evaluation, as it can capture the graded relevance values. Although the relevance is binary in our task setting, it is still a good indicator on how well the products are ranked relative to the ideal ranking.

\subsubsection{Baselines} \hfill\\
To verify the effectiveness of our model DualSI and answer the three research questions, we compare it with several baselines.
\begin{enumerate}
\item[-] {\bf QSBPS} \cite{zou2019learning}. The QSBPS model uses a Bayesian approach for sequentially asking users on their expected attributes. It represents the interactive search methods that queries user for explicit feedback. Specifically, the model directly queries the users on the expected presence of entities in the relevant product documents. It is based on the assumption that there is a set of candidate questions for each product to be asked.

\item[-] {\bf Qrec} \cite{Zou_2020_Qrec}. Qrec is an interactive recommendation method which recommend items to users by asking their preferences in a similar way as QSBPS. It is initialized offline with a matrix factorization model using the historical item ratings by users, followed by an online update of user and item latent factors based on user's answer. It learns to select the best question sequence to ask. Since there is no rating data in our dataset, we treat each purchase pair with a rating equal to 3. 
% Because majority of the sessions in the dataset are by anonymous users, each session is treated as conducted by a different user when applying Qrec to the dataset.

\item[-] {\bf SR-GNN} \cite{wu2019session}. SR-GNN is one of the state-of-the-art model for session-aware recommendation. It constructs the clicked items into graph to capture the complex transitions, and predicts the next item to be clicked. Although it is not specifically designed for search task, it can be used to represent those session-aware methods which leverage implicit feedback from user clicks. In our setting, the next item refers to the target item.

\item[-] {\bf SCEM} \cite{bi2020leverage}. The SCEM model makes use of the clicked item sequence for search result re-rank. Specifically, it leverages clicks within a query session, as implicit feedback, to represent users' hidden intents, which further act as the basis for re-ranking subsequent result pages for the query. We use it together with SR-GNN as baselines representing the product search methods that solely use click-stream data.

\item[-] {\bf DualSI} variations. The first variant DualSI\_randomQ randomly selects attributes to ask on top of our general framework. Similarly, DualSI\_randomA randomly answers the questions asked. In DualSI\_popular, questions are generated based on the most frequently appeared attributes in the clicked items.
\end{enumerate}

\subsection{Main Results {\bf (RQ1)}} 
As shown in Table \ref{table:mainResult}, our method largely outperforms the baselines across all metrics. When asking one or two questions are allowed, the proposed DualSI method outperforms QSBPS and Qrec by large margins. For example, it obtains 45.96\% and 33.96\% performance gain comparing with QSBPS regarding MRR respectively. It also outperforms the SR-GNN and SCEM which only considers user click-stream data. We show that making use of the implicit session feedback to model user interests while allowing question asking is indeed an effective way to assist the users. This addresses the {\bf RQ1}.

Besides, asking two questions/attributes helps to gather more information from the users, hence the performance is better. This is as expected for both the QSBPS and Qrec baselines, and the proposed model and its variations. 
% It can be seen that the Qrec model performs worse than the rest by a large percentage. This is expected as it is a method designed for recommendation task and the search query information is not used. In addition, there is no actual rating data for it to do initialization to its full extent.

The DualSI\_popular achieves relatively good performance as compared to other variations. It uses the most frequently appeared attributes among the clicked items for question generation. This result suggests that the key attributes are usually unchanged though the click path and they play an essential role on critical question asking. The interpretation of the other variations will be discussed in the next subsection.

We also observe that when no question asking is allowed, our method performs slightly worse than SCEM. This might be due to the fact that the Ranking Net of DualSI shares encoder with the Selection Net while the later one is trained to select critical questions. When no question is asked, the encoder part deviates from its original learning purpose and hence affect the ranking results. As for SR-GNN, it performs the worst. This might be because SR-GNN is not specifically designed for search task and does not take query into consideration. Moreover, it does not take any textual information of the items.
%SCEM encodes user embeddings and shares them across search sessions which helps to boost the ranking performance. While the proposed DualSI did not do this as we hope to avoid privacy issues.

\subsection{Qualitative Analysis}
\subsubsection{The Effect of Questions Asked {\bf (RQ2)}} \hfill\\
Two model variants, DualSI\_randomQ and DualSI\_randomA, are explored to answer {\bf RQ2}. We use the same training settings for each method, keeping all the components the same except the question part.
As it is shown in Table \ref{table:mainResult}, the experiment results on randomly generating questions (DualSI\_randomQ) are similar to that of SCEM and DualSI (No Question Asking) which both make use of item click data only. This is generally to be expected. Besides, since the questions are randomly generated, the number of attributes being asked does not matter too much and there is limited performance difference on asking one question or two questions. The large performance gap between this variant and our original model suggests the effectiveness of the questions selected. It validates our design of the utility score and the effectiveness of the transformer-based Selection Net structure.

The other variant DualSI\_randomA randomly generates answers to the questions asked. As our Selection Net guarantees the quality of the questions asked, randomly answering `Yes' or `No' still maintains the performance at an acceptable level. This might be due to the fact that randomly answer the questions may still obtain an approximately 50\% chance to guess it right. Therefore, the result obtained is better than that of DualSI\_randomQ. It is interesting to note that, under the variant DualSI\_randomA, the performance on asking 2 attributes performs worse than asking 1 attribute. After carefully inspecting the experimental instances, we suspect that it is probably because of the increased noise introduced when randomly answering the acceptance/rejection on more attributes. From a different angle, this actually suggests the importance of gathering the correct answers from the users to directly know their preferences. This would increase the chance of getting the correct target item with less operations and save users' time.

\subsubsection{The Position for Prompting  Questions {\bf (RQ3)}} \hfill\\
We also conducted experiments on promoting question at different positions of the user's browsing path. The experimental results are presented in Table \ref{table:ablation}. On one hand, by asking at a relatively later position in the browsing path, the performance of the model increases accordingly. This aligns with our expectation, as seeing more clicked items would provide a more representative context and assist on asking more critical questions accurately. On the other hand, despite of asking the questions early (1-3 item clicks before target item), our model still outperforms the baselines. This suggests that our method can help to shorten the user search sessions. In real application, it is possible to find a good balance between ranking performance and session shortening.

\section{Conclusion}
In this work, we proposed a dual-learning model {\bf DualSI} which consists of a Selection Net and a dual Ranking Net. It leverages both implicit user feedback from user's click-stream data and explicit user feedback by asking critical questions. Note that existing product search methods either cannot capture user's current detailed interest or require intensive user involvement for clarifications while most users do not have an exact item in mind at the initial stage. It is more appropriate to allow users to browse first and then clarify with him/her on the target item. Hence, we introduced the innovative idea of utility score in Selection Net which helps to select the most critical questions to ask by learning from the clicked items. After incorporating user's answer on the question, the dual Ranking Net then ranks the items. According to the experiment results on the public dataset, our model largely outperformed the state-of-the-art baselines. 

In the future work, we plan to make use of the rejected entities also for performance improvement. Moreover, instead of promoting questions at a fixed position, dynamically prompting questions when needed could be an interesting direction to continue exploring.

\section*{Acknowledgement}
This research is supported by Shopee Singapore Private Limited, Singapore Economic Development Board and Sea-NExT Joint Lab.

\balance
\bibliographystyle{ACM-Reference-Format}
\bibliography{reference}

\end{document}